\documentstyle[twoside,fleqn,espcrc2]{article}

\newcommand{\be}{\begin{equation}}
\newcommand{\ee}{\end{equation}}
\newcommand{\bea}{\begin{eqnarray}}
\newcommand{\eea}{\end{eqnarray}}
\newcommand{\no}{\noindent}
\newcommand{\un}{\underline}

\def\NPB{{\it Nucl.\ Phys.\ }{\bf  B}}
\def\NPBPS{{\it Nucl.\ Phys.}\ [Proc.\ Suppl.]\ }
\def\PLB{{\it Phys.\ Lett.\ }{\bf  B}}

\def\RMP{{\it Rev.\ Mod.\ Phys.}}
\def\PRD{{\it Phys.\ Rev.\ }{\bf  D}}

\title{QCD topology using 
 scale controlled cooling:\\
Densities and cooling invariant observables}
\author{
Ion-Olimpiu Stamatescu$^{1}$ 
and Christian Weiss$^{2}$\\
{\small $^1$Inst. Theor. Physik, Univ. Heidelberg and FESt, 
Heidelberg, Germany}\\
{\small $^2$Inst. Theor. Physik II, Ruhr-Universit\"at Bochum, Germany}
}

\date{}

\begin{document}
\maketitle



\no ABSTRACT: We aim at reducing the uncertainties inherent in the 
analysis of the topological structure by
using scale controlled smoothing and
 observables independent on the ``microscopic" description of the
instanton ensemble. \medskip

\no Investigating the role of instantons in the QCD vacuum by 
lattice simulations is a  challenging problem. Instantons are involved
in the $U_{\rm A}(1)$ symmetry breaking 
(cf the Witten - Veneziano formula  which relates quenched topological
susceptibility and $\eta'$ mass) and in chiral symmetry breaking 
(via zero modes of the Dirac operator), and they   lead to
 dynamical effects at intermediate distances.
Global properties like susceptibility and charge distributions
afford direct tests, while
the features of the local structure 
(ensemble, size and distance distributions)
are ingredients for dynamical models \cite{Scha}. To recall, an ${\bf R}^4$ (anti-)instanton (A)I  of size $\rho$ located at $x=0$ has 
charge (q) and action (s) densities:
\vspace{-0.2cm}
\bea
q(x)=\frac{1}{8 \pi^2}F^*F =
{6\,Q\over { \pi^2\rho^4}} \left[ 1 + 
\sum_{\mu=1}^4\left( {{x_{\mu}} \over {\rho}}\right)^2
\right]^{-4} \\ 
Q=\int d^4x\,q(x),\ \ 
S=\int d^4x\,s(x) = |Q|\,S_0 \nonumber \\
s(x)=S_0\,|q(x)|, \ \ S_0=8 \,\pi^2,\ \ 
Q= \pm\,{\rm integer}.\nonumber
\eea
 ``Superpositions" of $N$ 
I's {\it or}  A's lead to higher minima of the action:
 $S = N S_0$. However pairs I-A are not minima and, e.\,g., 
under unrestricted cooling they decay, the sooner the 
smaller their action 
  $S^{\rm IA}=2\,S_0 -S_{\rm int}^{\rm IA} < 2\,S_0$ is,
where $S_{\rm int}^{\rm IA}$ depends in particular  on  the ``overlap" $\omega=(\rho_{\rm I}+ \rho_{\rm A})/ d_{\rm IA}$, $d_{\rm IA}$ being the
I - A distance.\medskip

\no In analyzing the topology by
lattice methods we need to deal with the roughness of
the Monte Carlo configurations at short scales 
and to  identify  the physically relevant topological structure.
UV lattice
artifacts (dislocations) and close I-A pairs, indistinguishable from short 
range density 
fluctuations, have small action and
  can be
easily produced.
To smooth out in a 
controlled way high frequency fluctuations in Monte Carlo configurations
the method of {\it Restricted Improved Cooling} (RIC) 
has been developed in \cite{ric}
as a {\it gauge invariant low pass filter}. 
RIC  introduces a parameter 
 directly related to a physical scale above which fluctuations
will be  preserved. Since it uses an ``improved" action
RIC has  as fixed point non-trivial
classical configurations. 
But since it preserves
all structures above the chosen scale it also retains, e.\,g.,
 I-A pairs with
overlap smaller than a threshold depending on the 
smoothing scale. 
 
 The identification of the uncovered topological structure
poses special problems. Usually a ``microscopic" description in terms of
I's and A's is attempted. This, however,
 becomes increasingly ill defined at small scales,  particularly
 if close pairs abound. 
Therefore it is interesting to avoid the necessity of such a
description and obtain the phenomenological parameters appearing 
in the instanton models from observables which can be directly
measured on the lattice.

We here suggest to study properties of instantons in lattice simulations
using  observables  objectively
defined in continuum QCD. Rather than investigating instantons 
microscopically by inspecting lattice field configurations, we shall 
consider ``macroscopic" observables which in a pure instanton vacuum
can directly be related to properties of the instanton
ensemble. One such class of quantities are ratios of VEV's
of chirally odd operators, which are purely non-perturbative 
quantities, and (at least in a dilute instanton vacuum)
are
independent on the instanton density and can directly be related to 
moments of the instanton size distribution. Assuming the latter
to be a well defined property of the topological structure,  we expect these 
ratios to remain approximately constant under cooling, at least after 
achieving a
 certain, minimal degree of smoothing. Using RIC we 
can rephrase this question in terms of the physical 
scale of the relevant topological fluctuations.\medskip

\no {\bf Scale Controlled Smoothing.}   
 Cooling \cite{tep0} is an 
iterative, local minimization of the action, which 
proceeds sweep-wise and can be defined to 
converge onto (non-trivial) classical configurations. 
It acts  as a diffusion process with the length scale of smoothing 
growing like the square root of the number of iterations
\cite{tepe1}.

Restricted Improved Cooling 
(RIC) \cite{ric} is a scale controlled 
smoothing procedure involving two ingredients:

\no a) - {\it Improved} minimization action with practically scale
invariant instanton solutions, and

\no b) - {\it Restriction} of cooling to allow a certain amount of (Euclidean)
``energy" {\un {above}} the minimum, homogeneously distributed over 
the lattice.

\begin{figure}[htb]
\vspace{4.5cm}
\includegraphics{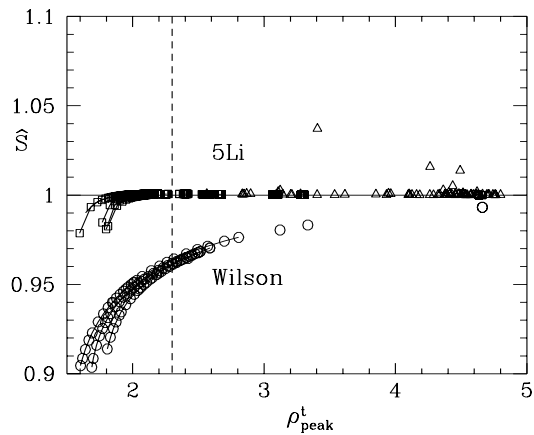}
\caption
{Improved (5Li) and Wilson action  vs instanton size 
given in lattice units - these are SU(2) results, but they
are representative also
 for SU(3).}
\label{f.act}
\vspace{-0.5cm}
\end{figure}

Our {\it improved} action \cite{ric}, \cite{mnpnp}
is correct to order ${\cal O}(a^6)$ and is completely flat for 
instanton sizes larger than $\rho_0 \sim 2.3a$, below which it drops
- see Fig. \ref{f.act}. 
It then follows from a) that:
 
\no - Below the ``dislocation threshold"
$\rho_0$ short range topological structure is eliminated.
 Note that $\rho_0 \rightarrow 0$ in  continuum.

\no - Above  $\rho_0$, 
instantons are stable to
 cooling.

\no - The correspondingly
{\it improved charge density} 
leads to a charge approaching an integer already after a few cooling
sweeps and stable thereafter.

 The {\it restriction} b) is introduced as the following modification
of the local updating rule for a link:
\be 
U \rightarrow V \ \ \, {\rm \un {iff}} \ \ \,
\Delta\equiv {\rm Tr}(WW^{\dagger}
-(U W^{\dagger})^2)\,\geq\,\delta^2 
\label{cond0}
\ee
\no where $W$ is the staple connected to $U$ and $V$ the group projection
of $W$ (for SU(2) $V=W/||W||$). It naturally leads to gradual saturation
and eventual stop of cooling after a number of sweeps depending on $\delta$. RIC thus acts as a frequency filter for 
the field fluctuations. Since $\Delta$ is the 
square of the lattice equations of motion  it has a continuum limit
and therefore the cooling parameter $\delta$ can be related to a physical scale. 

\begin{table}[htb]
\vspace{-0.5cm}
\begin{center}
\label{t.sat}
\begin{tabular}{|c|c|c|c|c|c|}
\hline
$\delta$&$r_c$& $\chi^{1/4}$&Sat. &$Plaq.$&$|P|$\\
 fm$^{-3}$& fm&MeV&sw.&&\\ 
\hline
\hline
\multicolumn{2}{|c|}{no cooling}&144(2)&0&.5754&.01\\ \hline
411.3  & .27(5)& 176(3)&8  &.9846&.04\\ \hline
290.8 & .34(6)&178(3)&16&.9888&.04  \\ \hline
145.4 & .41(7)& 181(3)& 30&.9939&.05  \\ \hline
51.41& .67(9)& 183(3)&67&.9976&.07\\ \hline
\end{tabular}
\label{t.ric}
\vspace{0.3cm}
\caption[]{RIC results for SU(3).}
\end{center}
\vspace{-0.3cm}
\end{table}

\begin{figure}[htb]
\vspace{4cm}
\includegraphics{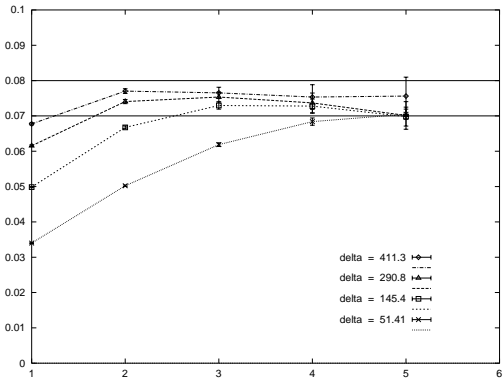}
\caption
{Mass gap $M(t) =\sigma\, L$  vs lattice distance 
$t$ ($L$: spatial lattice size). 
The horizontal band indicates 
standard results for this lattice. }
\label{f.sig}
\vspace{-0.3cm}
\end{figure}

\begin{figure}[htb]
\vspace{4cm}
\includegraphics{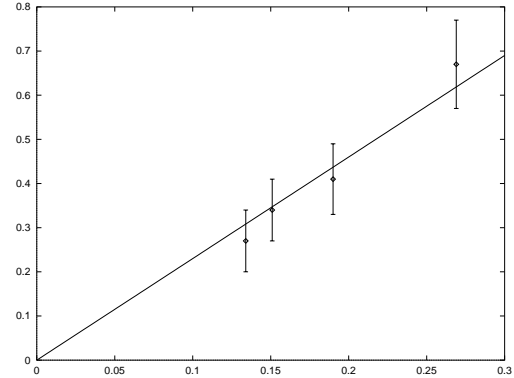} 
\caption
{ 
RIC smoothing scale $r_c(\delta)$ vs $\delta^{-1/3}$. }
\label{f.rc3}
\vspace{-0.3cm}
\end{figure}

 For this we calculate
the string tension $\sigma$ from correlations of spatial Polyakov loops at
separation $t$ in time, 
measured on configurations RI-cooled with $\delta$. We determine
 $r_c(\delta)$ as the the minimal
distance $t$  at which the string tension is recovered, taking the first
 cooling curve
as reference.
 See Fig. \ref{f.sig}; for details and for
SU(2) results see \cite{ric}.  The data presented here concern SU(3)
and have been obtained on
a $12^4$, pbc lattice at $\beta=5.85$ ($a=0.135\,$fm), 
using 900 configurations separated
by 600 sweeps (after 50000 thermalization sweeps). The line in Fig. \ref{f.rc3}
corresponds to $r_c(\delta) \approx 2.3 \ \delta^{-1/3}$.

\begin{figure}[htb]
\vspace{3cm}
\includegraphics{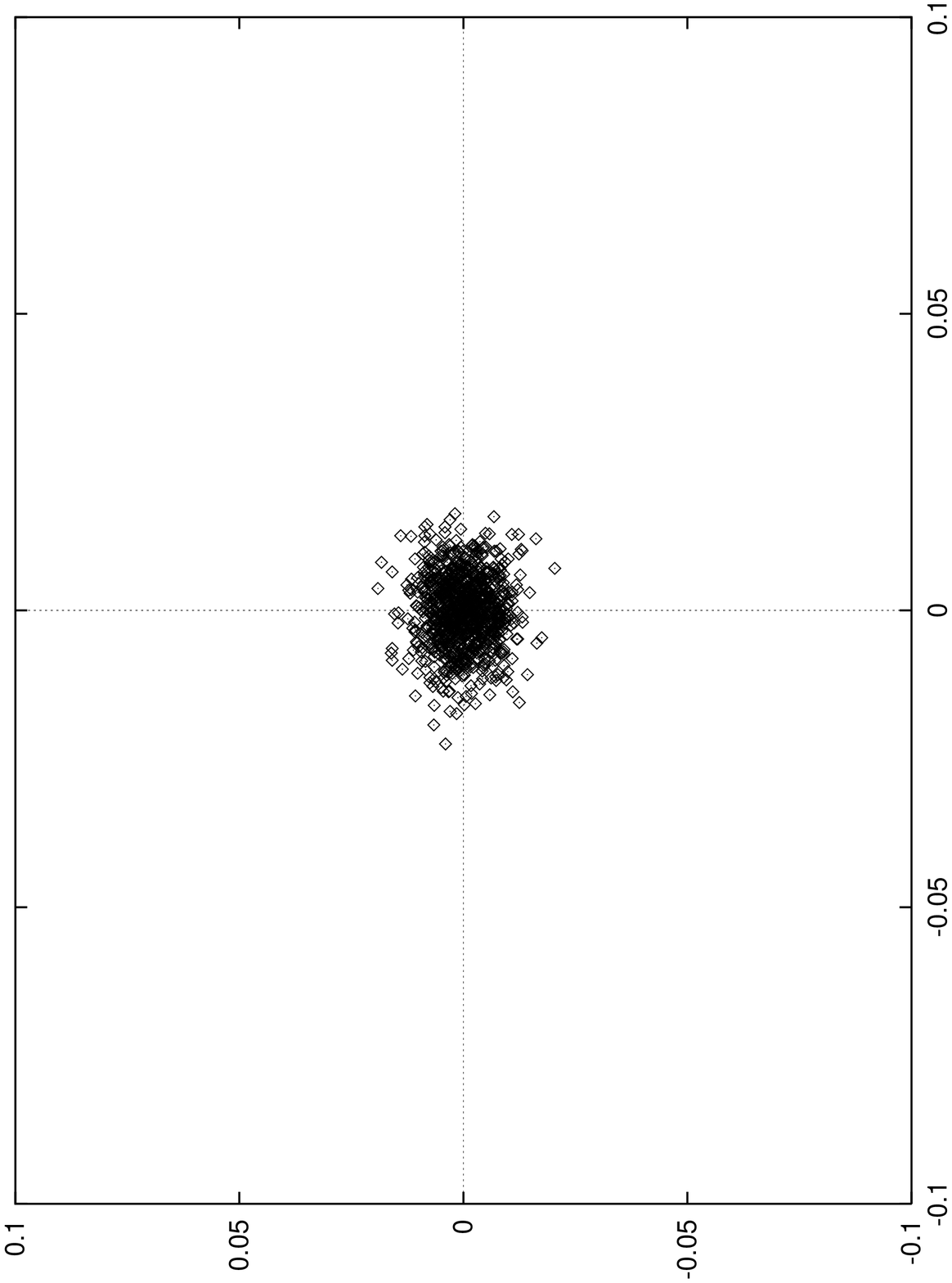}
\includegraphics{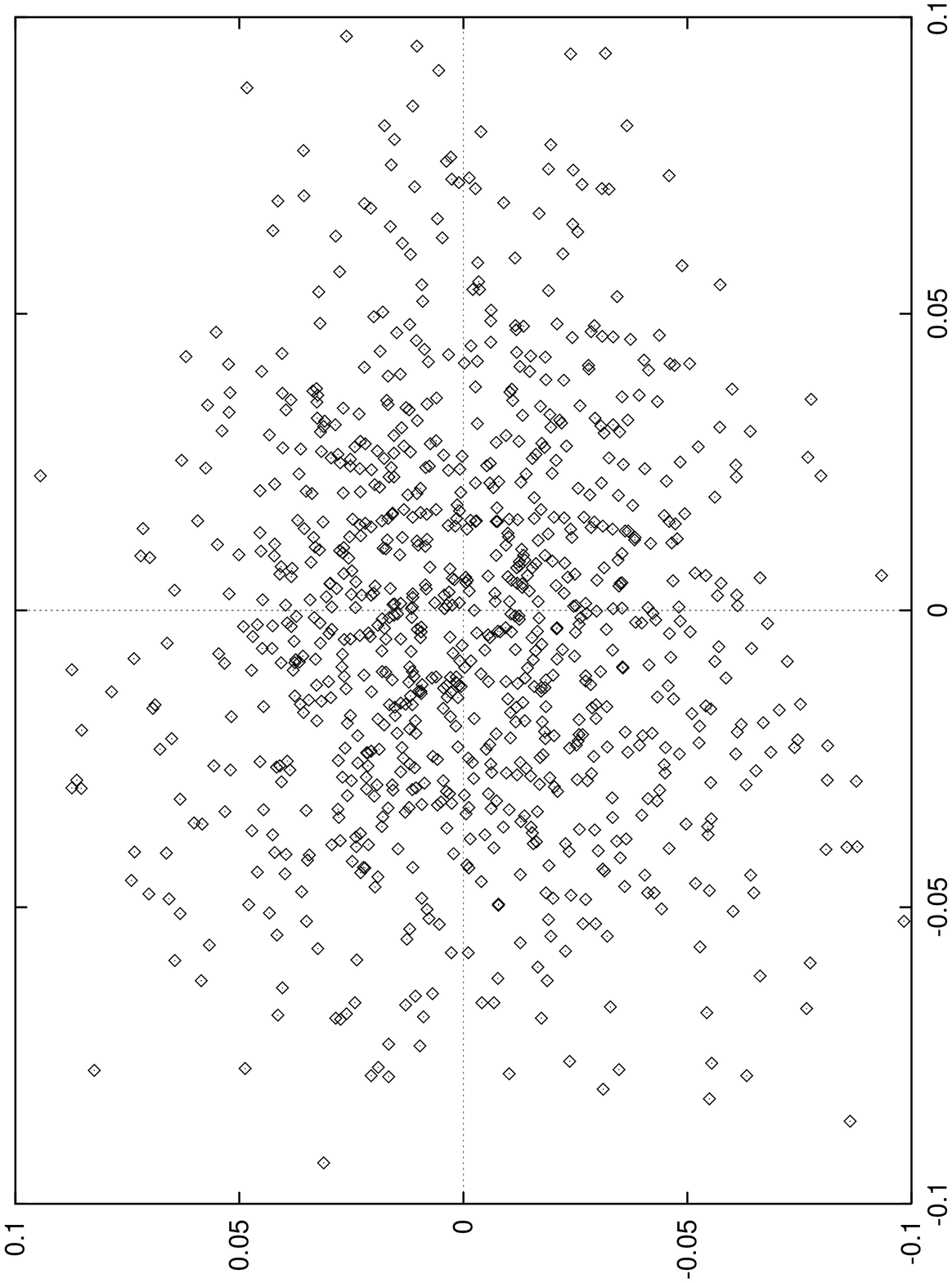}
\caption{Polyakov loop distribution in complex plane. Left: no cooling,
right: RIC ($\delta=290.8$).
}
\label{f.pol}
\vspace{-0.3cm}
\end{figure}

The restriction b) does not affect significantly the behavior of 
single instantons under RIC but does affect the behavior of I-A pairs. 
There is a well defined relation between $\Delta$
and the I-A ``interaction" $S_{\rm int}^{\rm IA}=16\pi^2-S^{IA}$ \cite{ric}.
 As a result
 RIC with given  $\delta$ {\it preserves} pairs depending on their interaction, hence 
 on their overlap. 
Generally, physics on scales larger than $r_c$ is
expected to remain unaffected by RIC. How RIC affects small and large scales
is also 
illustrated by the values of the Plaquette and Polyakov loop ($|P|$) in Table 1. The distribution of the latter after cooling stays compatible with confinement (see Fig. \ref{f.pol}). An extended topology analysis for SU(2) using RIC
 has been provided in \cite{ric}, here we add some further results for SU(3).

\begin{figure}[htb]
\vspace{4cm}
\includegraphics{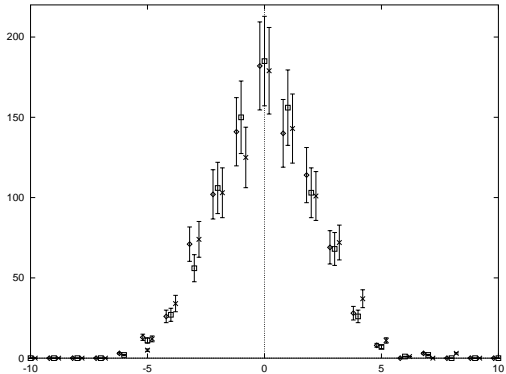}
\caption{Topological charge distribution from 
 RIC at $\delta=411.3$ (squares),
$290.8$ (diamonds, left shift) and $\delta=145.4$ (crosses, right shift). 
The tendency to depletion 
in the $|Q|=1$ sector with increasing smoothing is due to the periodic b.c.
}
\label{f.chd}
\vspace{-0.3cm}
\end{figure}

\begin{figure}[htb]
\vspace{3.5cm}
\includegraphics{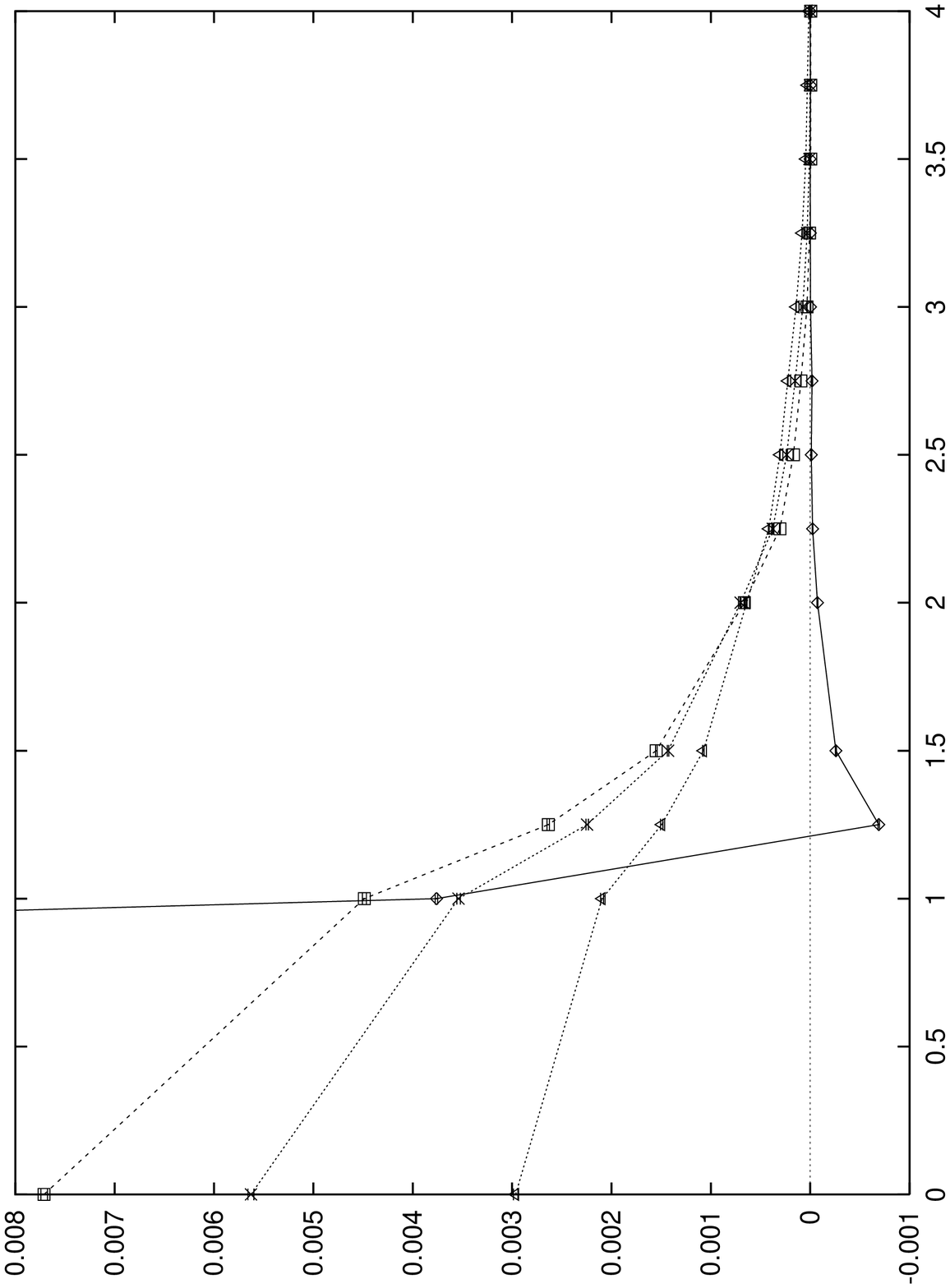}
\includegraphics{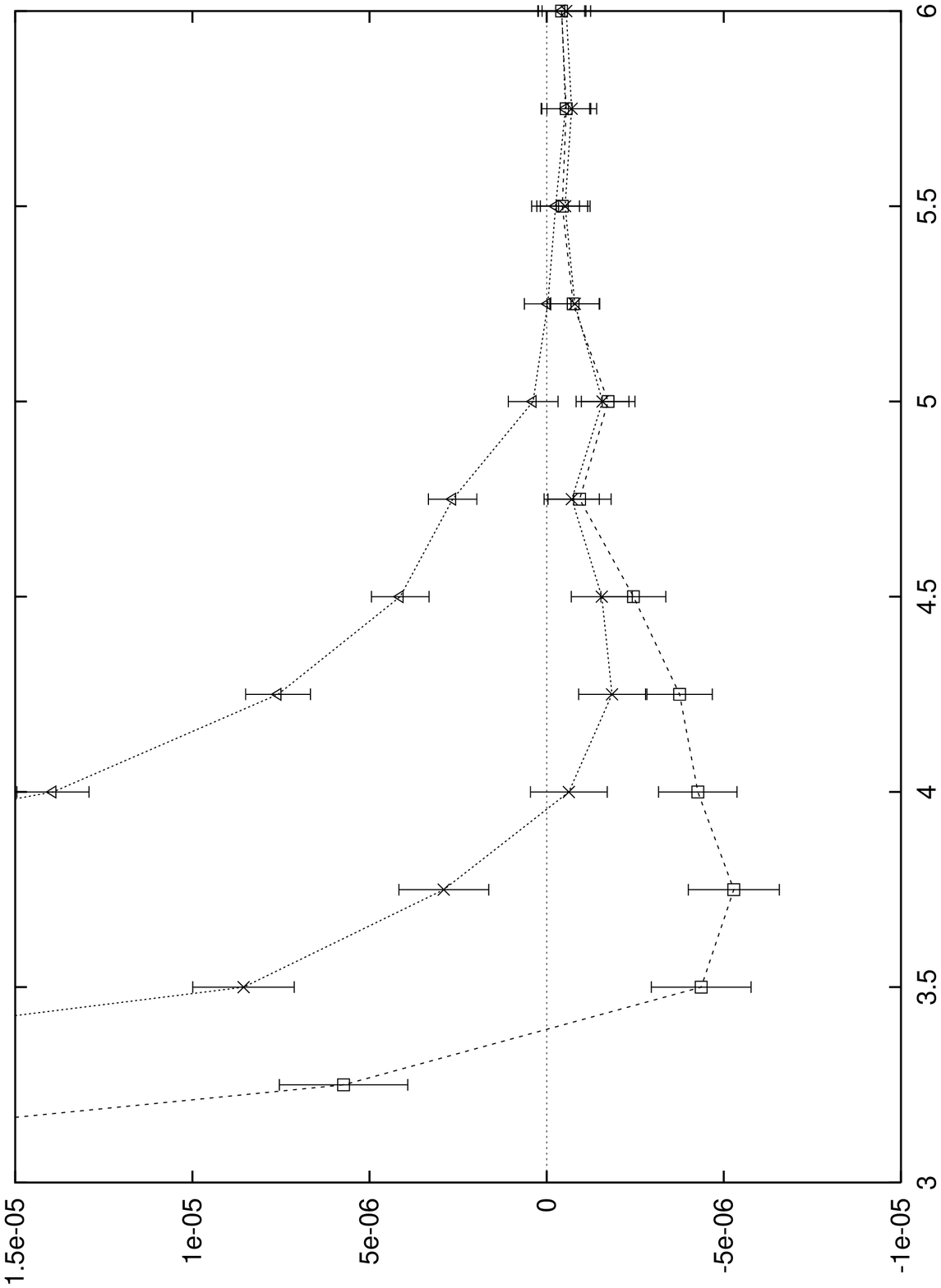}
\caption{Density correlations
vs lattice distance (binned) for no cooling and for RIC with the first three  
values for $\delta$ in Table 1 (curves from up to down). The right plot
shows the region of the minimum, enlarged, for the same cooled data 
(curves from left
to right).
}
\label{f.dens}
\vspace{-0.3cm}
\end{figure}

The topological charge distribution
 is Gaussian and is stable under cooling - see Fig. \ref{f.chd}.
In Table 1 we give the topological susceptibility of SU(3) for various $\delta$
(the phenomenological 
expectation is $\chi^{1/4} \sim 180\,$MeV). 
On Fig. \ref{f.dens} we plot the SU(3) density correlations 
$\langle q(x)\, q(0)\rangle$ at various values of $\delta$. Due to the 
improved charge density even non-cooled data 
can be obtained, they show however  strong UV renormalization effects.
Notice that
$\langle q(x)\,q(0)\rangle < 0$ for 
disjoint supports \cite{erha}. With progressing 
smoothing this effect disappears indicating gradual loss of
support properties. 

From the RIC analysis of SU(2) \cite{ric} it appeared 
 that one can speak of a typical size, but that a ``microscopic"
description of the instanton ensemble is  problematic
since topological excitations cannot be separated 
at small scales from short range fluctuations. We observed at small $r_c$
a strongly growing I(A) density, with more and more
peaks of alternating charge and increasingly  large overlap
 showing up at distances in the $r_c$ range. This may also affect the size determination and explain the diferences in the detail of the size
distribution found in the literature \cite{nege}.
To avoid the
uncertainties introduced by this situation in the determination of the
instanton size we look for 
an independent determination of the typical size using
 ``macroscopic" observables.
\medskip

\no{\bf ``Macroscopic" observables for topology.}
The relevance of instantons to chiral symmetry breaking is due to the
fact that a single $I$ ($A$) induces a localized (i.e., 
normalizable) zero mode of the Dirac operator of definite chirality:
\be
i \hat\nabla\,(A_{I (A)})\,\Phi_\pm (x)  =  0 ,\ \ 
\gamma_5\,\Phi_\pm (x)  =  \pm\, \Phi_\pm (x) .
\ee
Instanton models usually proceed from a simple picture
of the Yang--Mills vacuum as a dilute ``medium'' of $I$'s and 
$A's$:
\be
\bar\rho^4\,\Omega \ll 1 ,\ \ {\rm or} \ \ 
{\bar\rho}/{\bar R} \ll 1 
\ee
where $\bar\rho$ is the average size, $\bar R$ the average distance
and $\Omega = N/V$ the density of the instantons. 
The ratio ${\bar\rho}/{\bar R}$ can be used as
a small parameter to  classify non-perturbative 
effects generated by the medium of instantons
\cite{Shuryak:1982ff}, \cite{Diakonov:1984hh}. In this picture 
chiral symmetry breaking can be understood
as a collective effect involving all instantons in the ensemble 
 \cite{Banks:1980yr}, \cite{Diakonov:1986eg}
(this can be seen as a consequence of the cumulative effect of the 
chirally odd
't Hooft vertices of individual instantons, resulting in the
appearance of a dynamical quark mass). The
  chiral order parameter is then \cite{Diakonov:1986eg}:
\begin{equation}
\langle \bar\psi\,\psi \rangle  \sim  \bar\rho^{-3}
\left(\bar\rho^4\,\Omega\right)^{1/2}  \sim  
\bar\rho^{-3}({\bar \rho}/{\bar R})^{2}   
\end{equation}

Generally, the VEV's 
of any chirally odd operator can serve as  order parameter 
for chiral symmetry breaking. Since they are purely non-perturbative
quantities which acquire a non-zero value only because of the
spontaneous breaking of chiral symmetry they provide good
probes of the instanton effects. So, for instance, in an instanton
medium we have:
\begin{equation}
\langle \bar\psi\,{\cal F}[F]\,\Gamma\,\psi \rangle 
 \sim \bar\rho^{-d}\,\left(\bar\rho^4\,\Omega\right)^{1/2}
 \sim  \bar\rho^{-d}\,\left({\bar \rho}/{\bar R}\right)^{2}   
\end{equation}
where ${\cal F}[F]$ is a function of the gauge fields, 
 $\Gamma = 1,\, \gamma_5 , \,
\sigma_{\mu\nu}$ a chirally odd Dirac matrix
and $d$ is the mass dimension of the operator.
Note that while these VEV's depend differently
on the instanton size, they all  show the same dependence on the instanton 
density as the usual quark condensate. 
Therefore in {\it ratios} of such VEV's
the dependence on the instanton density cancels:
\begin{equation}
X  \equiv 
\langle \bar\psi\,{\cal F}[F]\,\Gamma\,\psi \rangle\, /\,
{\langle \bar\psi\,\psi \rangle}
 \sim  \bar\rho^{-d + 3} 
\label{R_generic}
\end{equation}
For example  one finds in the large--$N_c$ limit, where all
instantons are of size $\rho = \bar\rho$, \cite{Polyakov:1996kh}:
\begin{equation}
{\langle \bar\psi\,F_{\mu\nu}\,\sigma_{\mu\nu}\,\psi \rangle}\, /\,
{\langle \bar\psi \,\psi \rangle} =  4\,\bar\rho^{-2}
\label{sigma_operator}
\end{equation}
In the instanton vacuum  such ratios can thus directly be related
to properties of the instanton size distribution, with no reference
to the density.  (A non-zero density is needed only for numerator and
denominator to be non-zero individually.) Therefore we expect
quantities of the type (\ref{R_generic}) to be approximately 
``invariant'' under cooling.\medskip

\no We plan to measure the cooling behavior of  ratios of
the type (\ref{R_generic})
using RIC. 
Since we can control the smoothing scale we can  
test the dependence of our derivation on the assumption of diluteness
by observing the saturation of the ratios, and then obtain an
estimation of the typical instanton size independently on a 
microscopic description of the ensemble.
(A lattice simulation  for
 (\ref{sigma_operator}) has been performed with
limited statistics in \cite{Kremer:1987ve}.)\medskip

\no {\bf Acknowledgments}: IOS thankfully acknowledges  DFG support 
for attending the conference. The simulations are performed on the 
VPP computer of the University of Karlsruhe.

\end{document}